\documentclass[aps,prl,floatfix,superscriptaddress,twocolumn,tightenlines,amsmath,amssymb,nofootinbib]{revtex4-1}

\usepackage{graphics}
\usepackage{graphicx}
\usepackage{dcolumn}
\usepackage{bm}
\usepackage{amsmath,amssymb}
\usepackage{color}
\usepackage{soul,xcolor}
\setstcolor{red}

\makeatletter
\let\cat@comma@active\@empty
\makeatother

\begin{document}

\title{Ghost imaging with the human eye}

\author{Alessandro Boccolini}
\affiliation{Scottish Universities Physics Alliance (SUPA), School of Engineering $\&$ Physical Sciences, Heriot-Watt University, Edinburgh EH14 4AS, UK}

\author{Alessandro Fedrizzi}
\affiliation{Scottish Universities Physics Alliance (SUPA), School of Engineering $\&$ Physical Sciences, Heriot-Watt University, Edinburgh EH14 4AS, UK}

\author{Daniele Faccio}
\affiliation{School of Physics $\&$ Astronomy,  University of Glasgow, Glasgow G12 8QQ, UK}

\begin{abstract}
Computational ghost imaging relies on the decomposition of an image into patterns that are summed together with weights that measure the overlap of each pattern with the scene being imaged. These tasks rely on a computer.
Here we demonstrate that the computational integration can be performed directly with the human eye.  We use this human ghost imaging technique to evaluate the temporal response of the eye and establish the image persistence time to be around 20 ms followed by a further 20 ms exponential decay. These persistence times are in agreement with previous studies but can now potentially be extended to include a more precise characterisation of visual stimuli and provide a new experimental tool for the study of visual perception.
\end{abstract}

\maketitle

In ghost imaging, an object can be imaged despite never having interacted with the light recorded by the camera \cite{Padgett_review}. The object is illuminated with a structured light field, and the transmitted or reflected intensity is recorded with a ``bucket" detector with no spatial resolution. A second, spatially correlated light field is modulated with the recorded intensity pattern and projected onto a spatially-resolved detector such as a CCD camera which integrates over many frames to produce an image. 
Originally, ghost imaging was claimed to be a quantum effect, with spatial light correlations obtained from momentum-correlated photon-pair sources~\cite{pittman1995optical}. However it was soon realised that classical correlations, e.g. obtained from a laser beam split at a beamsplitter, achieved the same effect~\cite{gatti2004ghost}. In computational ghost imaging, only a single light field is required, and spatial light correlations are generated algorithmically~\cite{shapiro2008computational}. 

The simplest version of this approach involves raster scanning the scene point by point with a small laser spot (i.e. smaller than the features that we wish to resolve) and collecting the reflected intensity with a photodiode.
The scene is then reconstructed one `spot' at a time. Compressive sensing approaches are however, the preferred option: structured patterns---typically Hadamard patterns---of light are used to illuminate the scene~\cite{katz2009compressive}. The reflection or transmission intensities $A_i$ recorded by the bucket detector are used as multiplicative weights for each Hadamard pattern $H_i(x,y)$ before these are summed together to obtain an image of the object:
\begin{equation}\label{Hadamard}
O(x,y)=\sum^{N}_{i=1}A_iH_i(x,y)
\end{equation}
From~\eqref{Hadamard} we see that this protocol projects the object or scene onto a given (e.g. Hadamard) basis and finds the relative coefficients $A_i$ for each basis element $H_i$. This widely used approach can be extended to include full 3D information~\cite{howland2011photon}, or dynamic adjustment of the decomposition basis for foveated imaging~\cite{phillips2017adaptive}. It has also been applied to the temporal domain \cite{Genty,Devaux:16}. Ghost imaging allows imaging in situations where diffusion or scattering (between the object and the bucket detector) would otherwise compromise visibility \cite{ghost1,ghost2,Xu:15,Genty} and provides the capability to image objects at wavelengths for which single-photon cameras are not readily available \cite{Aspden:15}.
\begin{figure}[t!]
\centering{
\includegraphics[width = .8\columnwidth]{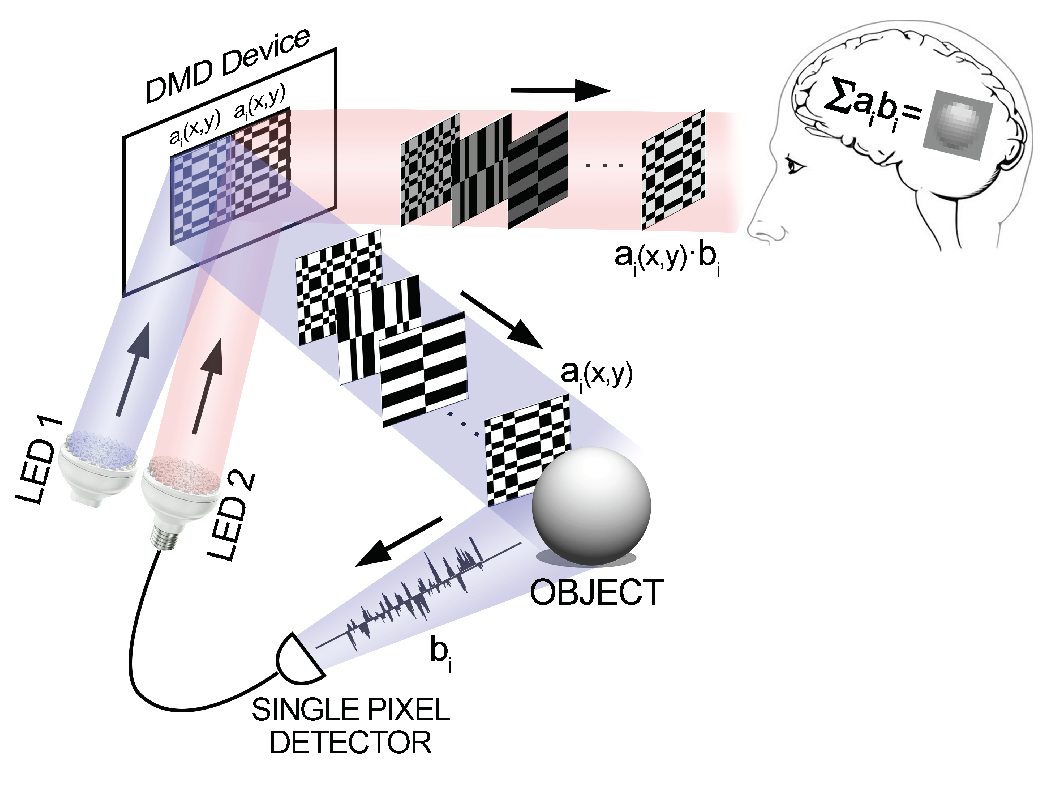}
\caption{Experimental ghost imaging with the human eye. LED1 illuminates the DMD which projects Hadamard patterns at 20 kHz onto an object. The reflected light is collected by a single-pixel detector. The output modulates the intensity of LED2 which also illuminates the DMD and is subject to the same patterns as LED1. The intensity-weighted Hadamard patterns are viewed on the DMD by eye or projected onto a screen. Human vision integrates over the  patterns when these are projected for much shorter durations than the eye's persistence time. As a result, although only black and white patterns are projected, the eye effectively perceives a ``ghost'' image of the object. }
\label{layout}}
\end{figure}

Apart from finding inspiration in human vision mechanisms to improve single-pixel imaging systems, one wonders if it is possible to implement ghost imaging directly in a biological system, using the human eye to replace parts of the imaging system or perform the required computational tasks. According to~\eqref{Hadamard}, once the coefficients $A_i$ have been measured, two computational tasks are required: (i) multiplication of each pattern with $A_i$; and (ii) integration over all weighted patterns, $A_iH_i(x,y)$.
In this work we confirm the capability of the human vision system to perform (ii), the integration of the weighted patterns. Humans can thus visualise a computational ghost image without any computer processing. We develop a macro-pixel approach to allow high resolution and detail in the perceived ghost image. Finally, by projecting the weighted patterns at increasing frame rates we can find a threshold below which a clear image is not distinguished. From this threshold we estimate the temporal response function of the human visual system.

{\bf{Theory.}} The problem of temporal persistence of images has been studied for a long time, see e.g. Ref.~\cite{Lollo}. Image persistence is used for example in cinematography where a series of still images are projected. If the frame rate exceeds the visual persistence time, then continuous and smooth movement is perceived: the transition between frames is blurred out by the convolution of the eye's response function and the projected image sequence.

Visual perception research has identified three mechanisms \cite{Col} that lead either individually or collectively to image persistence, namely: {\emph{neural persistence}}---activity in the visual system for a limited time  after the stimulus offset; {\emph{visible persistence}}---a visual stimulus that continues to be experienced for a limited time after its offset; and {\emph{iconic memory}}---the lingering of visual information that remains accessible for some time after the stimulus offset. More recently, attention has shifted towards the role of iconic memory and, in particular, the relation between attention and consciousness in perception \cite{Crick,Koch}. However, the overall process of visual perception and its underlying neurological mechanisms are still an object of debate (see e.g. Ref.~\cite{Block}).

In the following we will focus on the second of the three mechanisms outlined above, the visible persistence that allows us to perceive an image for a duration much longer than its projection time. Several studies have investigated image persistence time, using different methods that therefore also potentially measure different persistence mechanisms. One of the the main observations is that for very short ($<$100 ms) image projection (stimulus) times, the total visible persistence time of the image is of order 230 ms---regardless of the actual duration of the stimulus itself, as long as this remains below $\sim100$ ms \cite{Efron,Col}. These results relied on the measurement of synchronicity; the observers were asked to estimate the coincidence time between the offset of a stimulus with the onset of a second stimulus. Another series of measurements relied on asking observers to estimate the shape of forms that were broken into smaller parts that are then projected sequentially with varying permanence times  \cite{Lollo,Erik,Lof,Brockmole}. These studies led to the conclusion that complete images were observed through sequential projection of image parts as long as the temporal offsets where shorter than 50 ms, with a fast degradation and a breakdown complete image perception once temporal asynchronicity was above 100 ms.

We now build upon these findings and assume that the perception of a series of very short optical signals can be modelled as a temporal integration process with a specific response function, $R$, that consists of a short period of visual stimulus with constant amplitude, followed by a quick decay:
\begin{equation}\label{eye}
O(x,y,t)=\int R(t-t')I(x,y,t')dt'
\end{equation}
where $O(x,y,t)$ now is a time sequence of images, whereas $I(x,y,t')$ is the actual sequence of still images that are projected and $R(t-t')$ is the response function of the visual system. As explained above, we do not in general expect this to be a square function, i.e. a response that rapidly switches on and then, after a constant response level, suddenly switches off again (this is e.g. how a camera shutter would be described). For the moment, we do not need to make any specific estimate of the exact shape of $R$ and will return to this aspect later.

If the $N$ patterns in~\eqref{Hadamard} are provided sequentially (as opposed to all $N$ being available simultaneously), then we have a time dependent $I(x,y,t)=A(t)H(x,y,t)$ such that $A(t)=A_i$ and $H(x,y,t)=H_i(x,y)$ for $(i-1)\tau \leq t \leq i\tau$. If we detect this sequence of patterns with a true integrator, i.e. a detector that effectively has an infinitely long response time, then the sequence of $N$ patterns will provide the ghost image
\begin{equation}\label{eye_Hadamard1}
O(x,y)=\int A(t)H(x,y,t)dt.
\end{equation}
If instead we have a detector with instant response, we obtain 
\begin{equation}\label{eye_Hadamard2}
O(x,y,t)=\int \delta(t-t')A(t')H(x,y,t')dt',
\end{equation}
and the detector will just see the pattern sequence, and no ghost image. Equation~(\ref{eye}) constitutes an intermediate regime where the outcome depends on the projection rate or the relation between image persistence time $\tau$ and persistence time $\sigma$. For $\sigma\gg\tau$, a ghost image will be perceived, since~\eqref{eye_Hadamard1} well approximates~\eqref{eye}.
The visual system can perform the ``summation'' step in ghost imaging as long as the weighted Hadamard (or any other) set of patterns is projected at sufficient speed. Ideally, one will then cyclically repeat these projections to form a static image that is perceived for as long as the projections continue.

{\bf{Results.}} The experimental arrangement and procedure is show in Fig.~\ref{layout}. We performed a first set of experiments using a DMD that projected Hadamard patterns onto some object, with the reflected intensities detected by a bucket detector (a photodiode). The photodiode signal was fed into a light-emitting diode (LED) that illuminated a second DMD, synchronous to the first, and projecting the now intensity-weighted pattern onto a large screen. The test subjects viewed the screen and could vary the projected pattern frame rate. The DMD reaches 20 kHz frame rates and can thus project up to 200 Hadamard patterns within a 20 ms time window. At maximum frame rate, a clear image of the object (e.g. a hand) could be perceived by the viewer directly on the screen, without any intermediate computational processing. 
Slower projection rates quickly degrade the image visibility, resulting in `flickering' square patterns owing to the rapid succession of Hadamard patterns. This result, i.e. the fact that the human visual system can correctly reproduce a ghost image through summation of successive image frames was not a priori obvious due to the still unclear visual perception mechanisms and the fact that previous image integration studies used images with no overlapping areas of high luminosity \cite{Brockmole}.  

\begin{figure}[t!]
\centering{
\includegraphics[width = 3.5cm]{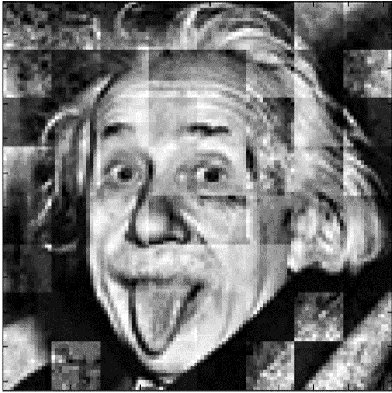}\hspace{.5cm}\includegraphics[width = 3.5cm]{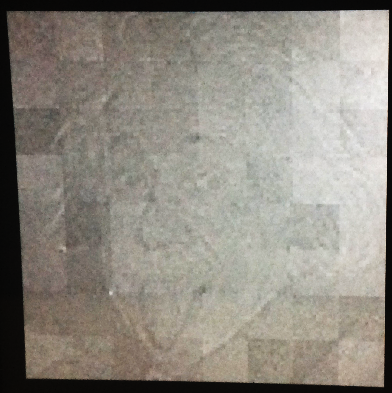}
\caption{Macro-pixel method for increased spatial resolution. The image on the left shows a computer simulation of the method with the original images divided in 64 macro-pixels, each of which is sampled with the same 256 Hadamard patterns. On the right we show the same image projected onto a screen and photographed with 50 ms exposure time.}
\label{einstein}}
\end{figure}

The ghost images described above were limited to a resolution of $15\times15$ pixels, since the patterns projected onto the screen are weighted by the photodetector reading and are therefore grayscale Hadamard patterns. Grayscale images are obtained from a DMD by ``dithering" the micro mirrors, i.e. 1 bit (black and white) images can be projected at 20 kHz but 4-bit images are limited to $\sim5$ kHz and a maximum of $\sim250$ grayscale patterns can be projected within a 50 ms time window. 

To increase the spatial resolution of the image we divided each image into a set of 8$\times$8 ``macro pixels''. Each of these was sampled with 256 Hadamard patterns, effectively increasing the resolution by a factor of 64. Figure~\ref{einstein} shows a $112\times112$ pixel image of Albert Einstein we used to simulate this method in practice. Each of the 64 macro pixels is sampled with the same 256 Hadamard patterns the DMD can project simultaneously. The left-hand image shows the computer-simulated process and reconstruction. The right-hand image shows the projected pattern on the viewing screen as photographed by a camera with 50 ms exposure time, providing a faithful rendition of what was perceived by a human observer, i.e. with temporal integration performed by the visual system. The system was tested by five different individuals and all agreed that they could clearly perceive the image and the similarity with the image shown in Fig.~\ref{einstein}. This demonstrates that it is possible to view relatively high-resolution images in a ghost-imaging setup using available projection technology and the human eye for image integration.

{\bf{Evaluation of visual response times. }} A key point in the results above is the ability for the DMD to project a sufficient number of frames within the persistence time so that these are temporally integrated and perceived as an image rather than a succession of black-and-white patterns. The DMD projection rate can be easily controlled and in doing so our test viewers agreed on a regime in which the projected images transitioned from black-and-white patterns to actual images. In the following we exploit this feature, in combination with the projection methodology based on macro-pixels, to evaluate the human visual response time. Our procedure is inspired by a classic motion persistence test, where a bright spot is moved quickly across a dark screen and is perceived by the eye to form a ``trail''. 

\begin{figure}[b!]
\centering{
\includegraphics[width = 7cm]{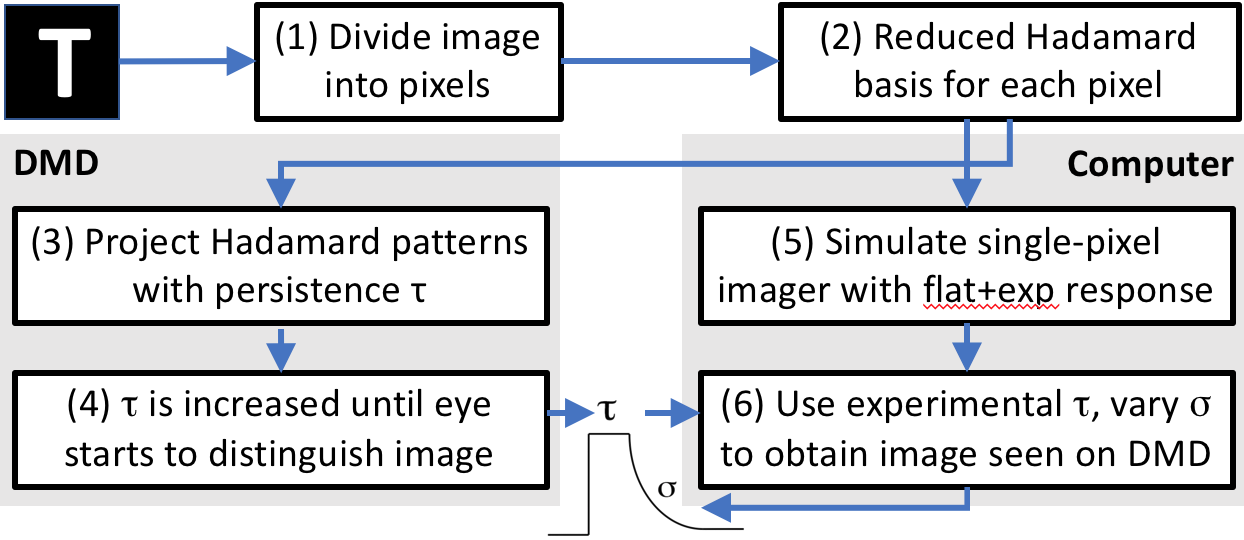}
\caption{Flow chart for estimating the response time of the human visual system.}
\label{diagram}}
\end{figure}

\begin{figure}[t!]
\centering{
\includegraphics[width = .8\columnwidth]{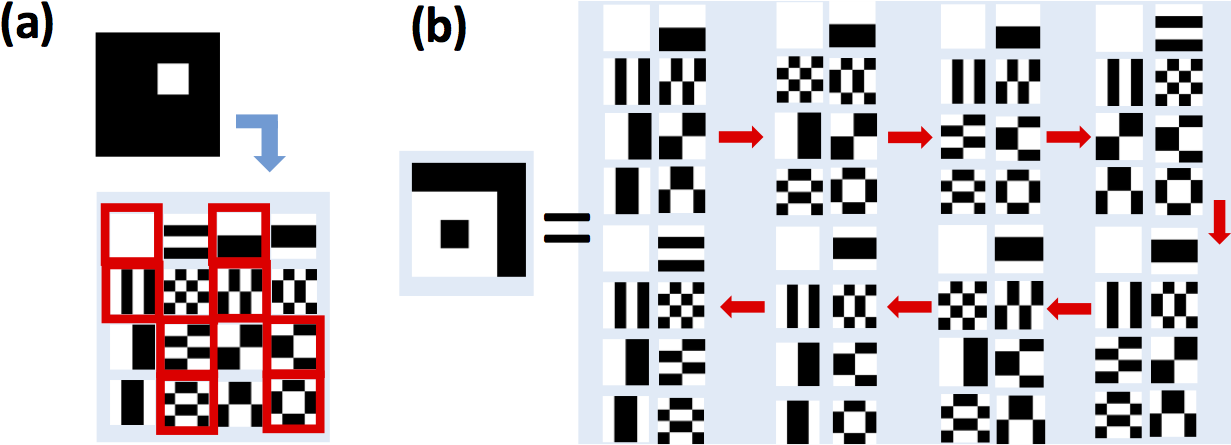}
\caption{(a) Decomposition of a single white pixel into a reduced set of Hadamard patterns (boxed in red). (b) Example of reduced set of Hadamard patterns for a 3x3 pixel ``zero''. The red arrows indicate the sequence with which each subset of Hadamard patterns (and hence the relative white pixels from the image) are projected by the DMD.}
\label{decomp}}
\end{figure}

Our approach is outlined in Fig.~\ref{diagram}. At step (1) we consider each of the white pixels individually and at step (2) we identify the Hadamard patterns required to reproduce that specific white pixel on a black background. The advantage in this approach is that we identify a reduced set of essential Hadamard patterns required to recreate each white pixel individually, minimising the number of patterns sequentially projected on the DMD. In step (3) we project each set of Hadamard patterns for each individual white pixel sequentially. Figure~\ref{decomp} shows an example for a 3$\times$3 pixel image of a ``zero''. Figure~\ref{decomp}(a) shows how a single pixel is decomposed into a reduced Hadamard set (boxed in red). Figure~\ref{decomp}(b) shows the complete set of reduced Hadamard patterns for the full image: the red arrows indicate the order in which each subset of Hadamard patterns (and hence also the relative white pixels from the image) are projected sequentially by the DMD. Each of these patterns is projected for a time $\tau$. The patterns are continuously looped to give the test viewer sufficient time to observe the projection on the screen. The experiment is then repeated, with a gradually decreasing $\tau$. We transition from a situation where for long $\tau$ individual Hadamard patterns are seen on the screen (and no image can be perceived), until the individual patterns can no longer be distinguished and a clear image emerges. The aim is to determine the ``threshold'' value for $\tau$ for which the patterns disappear and the images start to become partly visible. These tests were repeated with four test viewers and all reached the same conclusion that for $\tau\sim20$ ms, all of the projected patterns started to become visible.

Separately we simulate the experiment numerically, steps (5) and (6), by assuming that human visual perception has an initially flat response followed by an exponential decay (see the function in the Fig.~\ref{diagram} inset). We fixed the flat response time to the `threshold' image value for $\tau$ (20 ms). We therefore aim to determine the exponential decay time, $\sigma$ that is introduced to provide a connection with studies that found visible persistence times to gradually decay in time \cite{Col}. This mathematical model is by no means unique. However, a constant response followed by an exponential decay is the simplest option consistent with observations and, for our purpose, ideal to demonstrate the potential of ghost imaging for visual perception studies. We vary $\sigma$ in the simulations until a good match is found between the simulated pattern and the observed image.

\begin{figure}[t!]
\centering{
\includegraphics[width = .8\columnwidth]{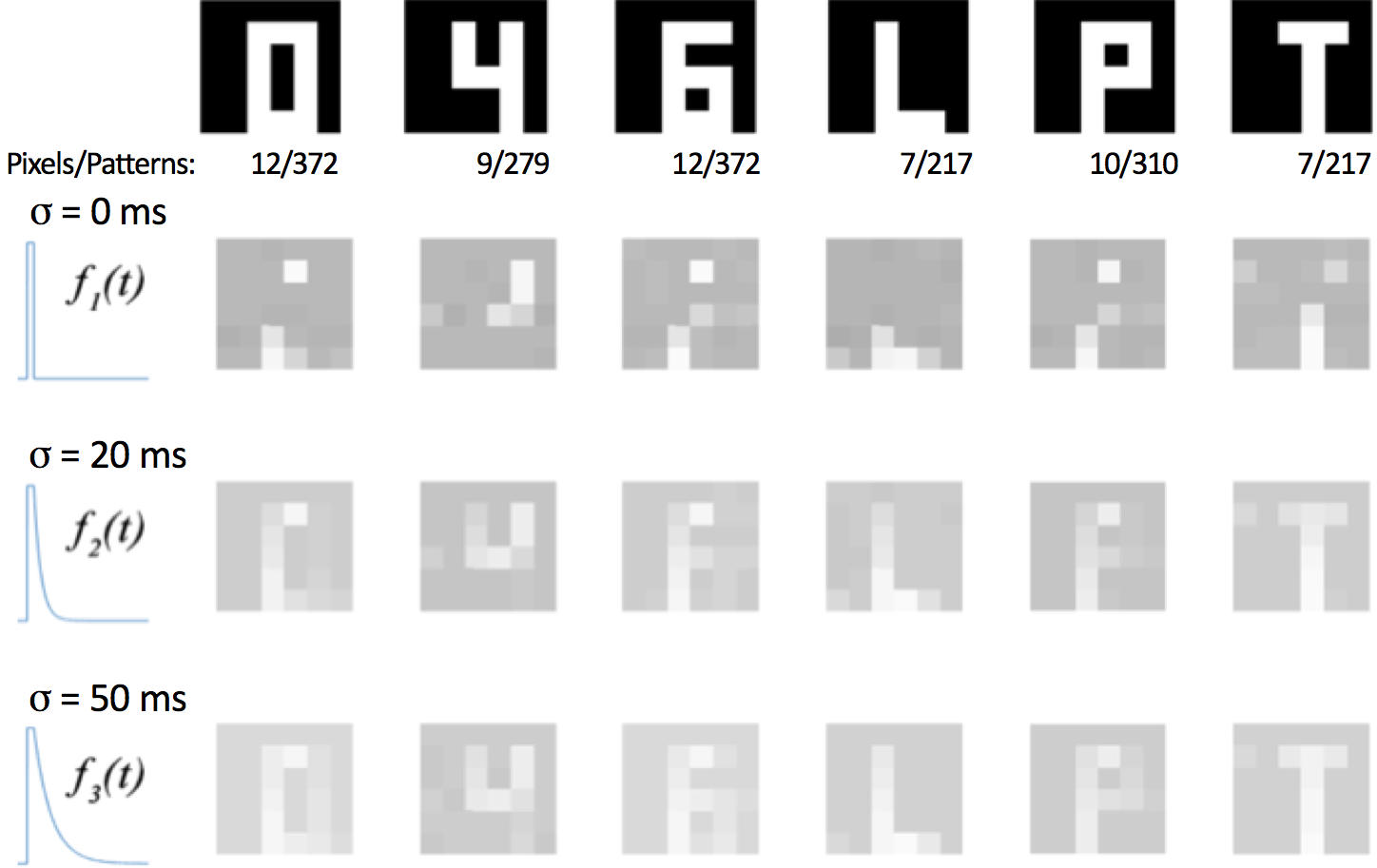}
\caption{Examples of simulated images used to evaluate visible persistence times. The top row shows the chose images (numbers and letters) together with the number of white pixels and the total number of Hadamard patterns required to represent the images following the recipe explained in Fig.~\ref{decomp}. The following rows show the reconstructed images with varying exponential decay times ($\sigma$) of the function $f(t)$ used to  weigh the projection of each individual Hadamard pattern.  }
\label{letters}}
\end{figure}

Figure~\ref{letters} shows examples from the experiments where we used $6\times6$-pixel images of the numbers ``0'', ``4'', ``6'' and letters ``L'', ``P'', ``T''. The top rows show the actual images used together with the number of white pixels and total number of Hadamard patterns projected. The successive rows show the simulated patterns for a range from $\sigma=0$ s (i.e. a square. step-like response, similar to a camera) to $\sigma=50$ ms. All four subjects aimed to match simulations and experiments for the case of image-formation threshold and found that the best match was obtained with $\sigma=20$ ms. For shorter $\sigma$ the image is hardly discernible (see e.g. second row, $\sigma=0$ ms) and for longer times the images appear complete and clearly visible (bottom row, $\sigma=50$ ms). Although the patterns were projected only for a fixed $\tau$, simulating the ghost image assuming a constant stimulus that lasts exactly $\tau$ cannot explain the observed projected images. An additional exponential decay time is required and using the concept of a `threshold' at which the images start to become discernible, can provide information about the decay rate.

{\bf{Discussion. }} We have demonstrated ghost imaging with the human eye, such that no computational steps are required for visualising an image that is just the projection of weighted (based on a proxy measurement) Hadamard patterns. We proposed a macro-pixel technique to increase the spatial resolution of the projected image. Ghost-imaging with the eye opens up a number of completely novel applications such as extending human vision into invisible wavelength regimes in real-time, bypassing intermediary screens or computational steps.

Perhaps even more interesting are the opportunities that ghost imaging offers for exploring neurological processes. Key to this approach is the substitution of computational steps with equivalent processes performed by the eye or visual cortex. Further investigation is required of how the perceived image depends on the specific shape of the visual response function together with an analysis of which visual mechanism is being probed. The persistence times we measured are closer to those obtained by performing `temporal integration' tests \cite{Lollo,Erik,Lof,Brockmole} rather than those based on estimates of synchronicity of multiple signals \cite{Efron,Col}. In our test, by comparing with a computer model we estimate the response time more precisely, splitting the response into a constant and decaying part.

A challenge could be to investigate if the visual system can perform the ``multiplication'' step in ghost imaging, which we obtained by weighting projected patterns with the bucket-detector intensities. This could be implemented using a VR system \cite{DMD-VR0,DMD-VR1} that simultaneously projects the (unweighted) Hadamard patterns and uniform grayscale weight images either onto both eyes separately or onto separate regions of the retina.

\section*{Funding Information}
DF and AF acknowledge support from the UK Engineering $\&$ Physical Sciences Research Council (EPSRC Grant No. EP/M01326X/1, EP/M006514/1, and EP/N002962/1, respectively).
%
%


\begin{thebibliography}{10}
\newcommand{\enquote}[1]{``#1''}

\bibitem{Padgett_review}
M.~J. Padgett and R.~W. Boyd, Philos. Trans. R. Soc. London A \textbf{375},
  2099 (2017).

\bibitem{pittman1995optical}
T.~Pittman, Y.~Shih, D.~Strekalov, and A.~Sergienko, Physical Review A
  \textbf{52}, R3429 (1995).

\bibitem{gatti2004ghost}
A.~Gatti, E.~Brambilla, M.~Bache, and L.~A. Lugiato, Physical review letters
  \textbf{93}, 093602 (2004).

\bibitem{shapiro2008computational}
J.~H. Shapiro, Physical Review A \textbf{78}, 061802 (2008).

\bibitem{katz2009compressive}
O.~Katz, Y.~Bromberg, and Y.~Silberberg, Applied Physics Letters \textbf{95},
  131110 (2009).

\bibitem{howland2011photon}
G.~A. Howland, P.~B. Dixon, and J.~C. Howell, Applied optics \textbf{50}, 5917
  (2011).

\bibitem{phillips2017adaptive}
D.~B. Phillips, M.-J. Sun, J.~M. Taylor, M.~P. Edgar, S.~M. Barnett, G.~M.
  Gibson, and M.~J. Padgett, Science advances \textbf{3}, e1601782 (2017).

\bibitem{Genty}
P.~Ryczkowski, M.~Barbier, A.~T. Friberg, J.~M. Dudley, and G.~Genty, Nat.
  Photon. \textbf{10}, 167 (2016).

\bibitem{Devaux:16}
F.~Devaux, P.-A. Moreau, S.~Denis, and E.~Lantz, Optica \textbf{3}, 698 (2016).

\bibitem{ghost1}
P.~Zhang, W.~Gong, X.~Shen, and S.~Han, Phys. Rev. A \textbf{82}, 033817
  (2010).

\bibitem{ghost2}
R.~E. Meyers, K.~S. Deacon, and Y.~Shih, Appl. Phys. Lett. \textbf{98}, 111115
  (2011).

\bibitem{Xu:15}
Y.-K. Xu, W.-T. Liu, E.-F. Zhang, Q.~Li, H.-Y. Dai, and P.-X. Chen, Opt.
  Express \textbf{23}, 32993 (2015).

\bibitem{Aspden:15}
R.~S. Aspden, N.~R. Gemmell, P.~A. Morris, D.~S. Tasca, L.~Mertens, M.~G.
  Tanner, R.~A. Kirkwood, A.~Ruggeri, A.~Tosi, R.~W. Boyd, G.~S. Buller, R.~H.
  Hadfield, and M.~J. Padgett, Optica \textbf{2}, 1049 (2015).

\bibitem{Lollo}
V.~D. Lollo, J. Exp. Psychol. \textbf{109}, 75 (1980).

\bibitem{Col}
M.~Coltheart, Perception {\&} Psychophysics \textbf{27}, 183 (1980).

\bibitem{Crick}
F.~Crick and C.~Koch, Nat. Neuroscience \textbf{6}, 119 (2003).

\bibitem{Koch}
C.~Koch and N.~Tsuchiya, Trends in Cognitive Sciences \textbf{11}, 16  (2007).

\bibitem{Block}
N.~Block, \enquote{The puzzle of perceptual precision,} in \enquote{Open MIND,}
   T.~K. Metzinger and J.~M. Windt, eds. (MIND Group, Frankfurt am Main, 2015),
  chap. 5(T).

\bibitem{Efron}
R.~Efron, Neuropsychologia \textbf{8}, 37 (1970).

\bibitem{Erik}
C.~Eriksen and J.~Collins, J. Exp.Psychology \textbf{74}, 476 (1967).

\bibitem{Lof}
G.~Loftus and D.~Irwin, Cognitive Psychology \textbf{35}, 135 (1998).

\bibitem{Brockmole}
J.~R. Brockmole, R.~F. Wang, and D.~E. Irwin, J. Exp. Psychology: Human
  Perception and Performance \textbf{28}, 315 (2002).

\bibitem{DMD-VR0}
P.~Lincoln, A.~Blate, M.~Singh, T.~Whitted, A.~State, A.~Lastra, and H.~Fuchs,
  IEEE Transactions on Visualization and Computer Graphics \textbf{22}, 1367
  (2016).

\bibitem{DMD-VR1}
P.~Lincoln, A.~Blate, M.~Singh, A.~State, M.~Whitton, T.~Whitted, and H.~Fuchs,
  \enquote{Scene-adaptive high dynamic range display for low latency augmented
  reality,} in \enquote{Proceedings of the 21st ACM SIGGRAPH Symposium on
  Interactive 3D Graphics and Games,}  (ACM, New York, NY, USA, 2017), I3D '17.
\end{thebibliography}


\end{document}